\documentclass[3p, twocolumn]{elsarticle}

\usepackage[boxed,algosection,linesnumbered]{algorithm2e}
\usepackage{epsfig}
\usepackage{bbm}
\usepackage{fullpage}
\usepackage{amssymb}
\usepackage{amsmath}
\usepackage{amscd}
\usepackage{amsthm}
\begin{document}
%
\title{Prepartition: Paradigm for the Load Balance of Virtual Machine Allocation in Data Centers}
\author[rvt]{Minxian Xu}
\ead{xmxyt900@gmail.com}
\author[rvt2]{Guangchun Luo\corref{cor1}}
\ead{gcluo@uestc.edu.cn}
\author[rvt2]{Ling Tian\corref{cor1}}
\ead{lingtian@uestc.edu.cn}
\author[rvt2]{Aiguo Chen\corref{cor1}}

\ead{agchen@uestc.edu.cn}
\author[rvt]{Yaqiu Jiang}
\ead{jiangyq\_uestc@hotmail.com}
\author[rvt]{Guozhong Li}
\ead{GuozhongLi@hotmail.com}
\author[rvt1]{Wenhong Tian\corref{cor1}  }
\ead{tian\_wenhong@uestc.edu.cn}

\cortext[cor1]{Corresponding author}
\address[rvt]{School of Information and Software Engineering, University of Electronic Science and Technology of China (UESTC), China}
\address[rvt1]{BigData Research Center, UESTC, China}
\address[rvt2]{School of Computer Science and  Engineering, UESTC, China}
\begin{abstract}
It is significant to apply load-balancing strategy to improve the performance and reliability of resource in data centers. One of the challenging scheduling problems in Cloud data centers is to take the allocation and migration of reconfigurable virtual machines (VMs) as well as the integrated features of hosting physical machines (PMs) into consideration. In the reservation model, the workload of data centers has fixed process interval characteristics. In general, load-balance scheduling is NP-hard problem as proved in many open literatures. Traditionally, for offline load balance without migration, one of the best approaches is LPT (Longest Process Time first), which is well known to have approximation ratio 4/3. With virtualization, reactive (post) migration of VMs after allocation is one popular way for load balance and traffic consolidation. However, reactive migration has difficulty to reach predefined load balance objectives, and may cause interruption and instability of service and other associated costs. In view of this, we propose a new paradigm, called Prepartition,  it proactively sets process-time bound for each request on each PM and prepares in advance to migrate VMs to achieve the predefined balance goal. Prepartition can reduce process time by preparing VM migration in advance and therefore reduce instability and achieve better load balance as desired. Trace-driven and synthetic simulation results show that Prepartition for offline scheduling has 10$\%$-20$\%$ better performance than the well known load balancing algorithms with regard to average utilization, imbalance degree, makespan  as well as capacity$\_$makespan. We also apply the Prepartition to online (PrepartitionOn) load balance and compare it with existing online scheduling algorithms, in which PrepartitionOn can  improve 8\%-20\% performance with regard to average CPU utilization, imbalance degree, makespan  as well as capacity$\_$makespan.
Both theoretical and experimental results are provided.
\end{abstract}

\maketitle

\section{Introduction}
In traditional data centers, applications are tied to specific physical servers that are often over-provisioned to deal with upper-bound workload. Such configuration makes data centers expensive to maintain with wasted energy and floor space, low resource utilization and significant management overhead. With virtualization technology, today's Cloud data centers become more flexible, secure and provide better support for on-demand allocating. The definition and model defined by this paper are aimed to be general enough to be used by a variety of Cloud providers and focus on the Infrastructure as a Service (IaaS). Cloud datacenters can be a distributed network in structure, containing many compute nodes (such as servers), storage nodes, and network devices. Each node is formed by a series of resources such as CPU, memory, and network bandwidth and so on, which are called multi-dimensional resources; each has its corresponding properties. Under virtualization, Cloud data centers should have ability to migrate an application from one set of resources to another in a non-disruptive manner. Such ability is essential in modern cloud computing infrastructure that aims to efficiently share and manage extremely large data centers. Reactive migration of VMs is widely proposed for load balance and traffic consolidation.

One key technology playing an important role in Cloud data centers is load balance scheduling. There are quite many load balance scheduling algorithms. Most of them are for traditional web servers but do not consider VM reservations with lifecycle characteristics. One of the challenging scheduling problems in Cloud data centers is to consider allocation and migration of reconfigurable VMs and integrated features of hosting PMs. The load balance problem for VM reservations considering lifecycle is as follows: given a set of $m$ identical machines (PMs) $PM_1, PM_2, \ldots, PM_m$ and a set of $n$ requests (VMs), each request [$s_i$, $f_i$, $d_i$], has a start-time ($s_i$), end-time ($f_i$) constraint and a capacity demand ($d_i$) from a PM, the objective of load balance is to assign each request to one of PMs so that the loads placed on all machines are balanced or the maximum load is minimized. This problem is not  well studied yet in the open literatures. The major contributions of this paper are:
\begin{itemize}
\item Providing a modeling approach to VM reservation scheduling with capacity sharing by modifying traditional interval scheduling problem and considering life cycles characteristics of both VMs and PMs.
\item Designing and implementing load balancing scheduling algorithms, called Prepartition for both offline and online scheduling which can prepare migration in advance and set process time bound for each VM on a PM.
\item Deriving computational complexity and quality analysis for both offline and online Prepartition.
\item Providing performance evaluation of multiple metrics such as average utilization, imbalance degree, makespan, time costs as well as capacity$\_$makespan by simulating different algorithms using trace-driven and synthetic data.
\end{itemize}
The remaining parts of this paper are organized as follows: Section 2 discusses the related work on load balance algorithms. Section 3 introduces problem formulation. Section 4 presents Prepartition algorithm in details as well as offline and online algorithms are described and compared. Performance evaluations of different scheduling algorithms are shown in section 5. Finally in section 6, a conclusion is given.
\section{Related works}
A large amount of work has been devoted to the schedule algorithms and can be mainly divided into two types: online load balance algorithms and offline ones. The major difference lies in that online schedulers only know current request and status of all PMs but offline schedulers know all the requests and status of all PMs.

Andre et al.\cite{IEEEhowto:Andre} discussed the detailed design of a data center.
Armbrust et al.\cite{IEEEhowto:Armbrust} summarized the key issues and solutions in Cloud computing. Foster et al.\cite{IEEEhowto:Foster} provided detailed comparison between Cloud computing and Grid computing.
Buyya et al.\cite{IEEEhowto:Buyya} introduced a way to model and simulated Cloud computing environments.
Wickremasinghe et al.\cite{IEEEhowto:Wickremasinghe} introduced three general scheduling algorithms for Cloud computing and provided simulation results. Wood et al.\cite{IEEEhowto:Wood} introduced techniques for virtual machine migration with spots and proposed a few reactive migration algorithms. Zhang\cite{IEEEhowto:Zhang1} compared major load balance scheduling algorithms for traditional Web servers.
Singh et al.\cite{IEEEhowto:Singh} proposed a novel load balance algorithm called VectorDot which deals with hierarchical and multi-dimensional resources constraints by considering both servers and storage in a Cloud. Arzuaga et al.\cite{IEEEhowto:Arzuaga} proposed a quantifying measure of load imbalance on virtualized enterprise servers considering reactive live VM migrations. Galloway et al. in \cite{IEEEhowto:Galloway} introduced an online greedy algorithm, in which PMs can be dynamic turned on and off but the life cycle of a VM is not considered. Gulati et al \cite{IEEEhowto:Gulati} presented challenge issues and Distributed Resource Scheduling (DRS) as a load balance scheduling for Cloud-scale resource management in VMware.

Tian et al.\cite{IEEEhowto:Tian1} provided a comparative study of major existing scheduling strategies and algorithms for Cloud data centers. Sun et al.\cite{IEEEhowto:Sun} presented a novel heuristic algorithm to improve integrated utilization considering multi-dimensional resource. Tian et al.\cite{IEEEhowto:Tian2} designed a toolkit for modeling and simulating VM allocation, \cite{IEEEhowto:Tian3}\cite{IEEEhowto:Tian4} introduced a dynamic load balance scheduling algorithm considering only current allocation period and multi-dimensional resource but without considering life-cycles of both VMs and PMs. Li et al.\cite{IEEEhowto:Li} proposed a cloud task scheduling policy based on ant colony optimization algorithm to balance the entire system and minimize the makespan of a given task set. Hu et al.\cite{IEEEhowto:Hu} stated an algorithm named Genetic, which can calculate the history data and current states to choose an allocation.

Most of existing research does not consider fixed interval constraints of VM allocation. Knauth et al.\cite{Knauth2012} introduced energy-efficient scheduling algorithms applying timed instances that have a priori specified reservation time of fixed length, these assumptions are also adopted in this paper. Most of existing research considers reactive VM migrations as a mean for load balance in data centers. To the best of our knowledge, proactive VM migration by pre-partition has not been studied yet in the open literatures. It is one of major objectives in this paper.

\section{Problem Formulation}
\subsection{Problem description and formulation}
In this paper we consider VMs reservation and model the VM allocations as a modified interval scheduling problem (MISP) with fixed processing time. More explanation and analysis about traditional interval scheduling problems with fixed processing time can be found in \cite{IEEEhowto:Kleinberg2} and references there in. We present a general formulation of modified interval-scheduling problem and evaluate its results compared to well-known existing algorithms.
There are following assumptions:\\
1) All data are deterministic and unless otherwise specified, the time is formatted in slotted windows. 
we partition the total time period [0, T] into slots with equal length $(s_0)$, the total number of slots is $k$=$T/s_0$. The start time $s_i$ and finish time $f_i$ are integer numbers of one slot. Then the interval of a request can be represented in slot format with (start-time, finish-time). For example, if $s_0$=5 minutes, an interval (3, 10) means that it has start time and finish time at the 3rd-slot and 10th-slot respectively. The actual duration of this request is (10-3)$\times$5=35 minutes.\\
2) For all VM reservations, there are no precedence constraints other than those implied by the start-time and finish-time.\\
3) The required capacity of each request is a positive real number between (0,1]. Notice that the capacity of a single physical machine is normalized to be 1 and the required capacity of a VM can be 1/8, 1/4 or 1/2 or other portions of the total capacity of a PM. This is consistent with widely adopted practice in Amazon EC2 \cite{IEEEhowto:Amazon} and \cite{Knauth2012}. \\

A few key definitions are explained as follows:

\textit{Definition 1. Traditional interval scheduling problem (TISP) with fixed processing time}:
A set of requests $\{$1, 2,$\ldots$, $n\}$ where the $i$-th request corresponds to an interval of time starting at $s_i$ and finishing at $f_i$, each request needs a capacity of 1, i.e. occupying the whole capacity of a machine during fixed processing time.

\textit{Definition 2. Interval scheduling with capacity sharing (ISWCS)}:
The only difference from TISP is that a resource (to be concrete, a PM) can be shared by different requests if the total capacity of all requests allocated on the single resource at any time does not surpass the total capacity that the resource can provide.

\textit{Definition 3. Sharing compatible intervals for ISWCS}:
A subset of intervals with total required capacity not surpass the total capacity of a PM at any time, therefore they can share the capacity of a PM.
In the literature, the makespan is used to measure the load balance, which is simply the maximum total load (processing time) on any machine. Traditionally, the makespan is the total length of the schedule.

In view of the problem in ISWCS for VM scheduling, we redefine the makespan as capacity$\_$makespan.

\textit{Definition 4. Capacity$\_$makespan of a PM $i$}: In any allocation of VM requests to PMs, let $A(i)$ denote the set of VM requests allocated to machine $PM_i$. Under this allocation, machine $PM_i$ will have total load equal to the sum of product of each required capacity and its duration (called Capacity$\_$makespan, i.e., CM for abbreviation in this paper), as follows:
\begin{equation}
CM_i=\sum_{j\in A(i)}d_j t_j
\end{equation}
where $d_j$ is the capacity requests of $VM_j$ from a PM and $t_j$ is the span of request $j$ (i.e., the length of processing time of request $j$).

Therefore, the goal of load balancing is to minimize the maximum load (capacity$\_$makespan) on any PM. Some other related metrics such as average utilization and makespan are also considered and will be explained in the following section.
Assuming there are $m$ PMs in data centers, the problem of ISWCS load balance in it therefore can be formulated as:
\textbf{
\begin{align}
& {Min_{1\le i\le m} ~~CM_i} \ \\
&\text{subject to 1). }~ \forall ~slot~s, \sum_{VM_j \in PM_i} d_{j} \leq 1 \\
& \text{~~~~ 2). } ~\forall~j, s_j~and~ e_j~are~fixed~by~reservation.
\end{align}
}
\noindent where $d_j$ is the capacity requirement of VM $j$ and the total capacity of a PM $i$ is normalized to 1. The condition 1) shows the sharing capacity constraint and condition, 2) is for the interval constraint of VM reservations.

\textit{ Theorem 1: The offline scheduling problem of finding an allocation of minimizing the makespan in general case is NP-complete.}
\\
The proof can be found in \cite{IEEEhowto:Tian4} and is omitted here.
\subsection{Metrics for ISWCS load balancing algorithm}
In this section, a few metrics closely related to ISWCS load balance problem will be presented. Some other metrics can be found in \cite{IEEEhowto:Tian4}.\\
1) PM resource: \\$PM_i(i,PCPU_i,PMem_i,PSto_i)$, $i$ is the index number of PM, $PCPU_i$,$PMem_i$,$PSto_i$ are the CPU, memory, storage capacity of that a PM can provide.\\
2) VM resource: \\$VM_j(j,VCPU_j,VMem_j,VSto_j,T_j^{start},T_j^{end})$, $j$ is the VM type ID, $VCPU_j,VMem_j,VSto_j$ are the CPU, memory, storage requirements of $VM_j$, $T_j^{start},T_j^{end}$ are the start time and end time, which are used to represent the life cycle of a VM.\\
3) Time slots: we consider a time span from 0 to $T$ be divided into slots with same length. The $n$ slots can be defined as $[(t_0, t_1),(t_1, t_2),\ldots,(t_{n-1},t_{n})]$, each time slot $T_k$ means the time span $(t_{k-1}, t_{k})$.\\
4) Average CPU utilization of $PM_i$ during slot 0 and $T_n$ is defined as:
\begin{equation}
PCPU_i^U=\frac{\sum_{k=0}^{n} (PCPU_i^{T_k}\times T_k)}{\sum_{k=0}^{n} T_k}
\end{equation}
where $PCPU_i^{T_k}$ is the average CPU utilization during slot $T_k$. Average memory utilization ($PMem_i^U$) and storage utilization ($PSto_i^U$) of both PMs can be computed in the same way. Similarly, average CPU (memory and storage) utilization of a VM can be computed.\\
5) Makespan: the total length of a schedule for a set of VM reservations, i. e., the difference between the start-time of the first job and the finishing time of the last job.\\
6) The capacity$\_$makespan (CM) of all PMs: can be formulated as:
\begin{align}
CM=& \max_i{(CM_i)}
\end{align}
\\
From these equations, we notice that life cycle and capacity sharing are two major differences from traditional metrics such as makespan which only considers process time (duration).
Traditionally Longest Process Time first (LPT) \cite{Graha} is widely used for load balance of offline multi-processor scheduling. Reactive (post) migration of VMs is another popular way of load-balancing. \textit{However, reactive migration has difficulty to reach predefined load balance objectives, and may cause interruption and instability of service and other associated costs}. By considering both fixed process intervals and capacity sharing properties in Cloud data centers, we propose  new offline and online algorithms  as follows.
\section{Prepartition Algorithm} \label{Model}
\subsection{Offline Prepartition Algorithm}
For a given set of VM reservations, let us consider there are $m$ PMs in a data center and denote OPT as the optimal solution for a given set of $J$ VM reservations. Firstly define
\begin{equation}
P_0=max\{max_{j=1}^{J} {CM_j},\frac{1}{m}\sum_{j=1}^{J} CM_j\}\leq OPT
\end{equation}
$P_0$ is a lower bound on OPT.
Algorithm 4.1 shows the pseudocodes of Prepartition algorithm. The algorithm firstly computes balance value by equation (7), defines partition value ($k$) and finds the length of each partition (i.e. $\lceil P_0/k\rceil$, which is the max time length a VM can continously run on a PM). For each request, Preparition equally partitions it into multiple $\lceil P_0/k\rceil$ subintervals if its CM is larger than $\lceil P_0/k\rceil$, and then finds a PM with the lowest average capacity$\_$makespan and available capacity, and updates the load on each PM. After all requests are allocated, the algorithm computes the capacity$\_$makespan of each PM and finds total partition (migration) numbers. For practice, the scheduler has to record all possible subintervals and their hosting PMs of each request so that migrations of VMs can be conducted in advance to reduce overheads.

\begin{algorithm}[!tb]
\SetArgSty{textnormal}
\caption{The pseudo codes of Offline Prepartition algorithm}\label{Prepartition }
\KwIn{VM requests indicated by their (required VM type IDs, start times, ending times, requested capacity), $CM_i$ is the capacity\_makespan of request $i$}
\KwOut{Assign a PM ID to all requests and their partitions}
Initialization: computing the bound $P_0$ value and set the partition value $k$;

\If{ $CM_i > P_0$ }
{divide it by $\lceil{P_0/k}\rceil$ subintervals equally and consider each subinterval as a new request}
Sort all intervals in decreasing order of $CMs$, break ties arbitrarity;

Let $I_1,I_2,\ldots,I_n$ denote the intervals in this order;

\ForAll{$j$ from $I$ to $n$} {\nllabel{Prepartition Outer Loop Begin Line}
Pick up the VM with the earliest start time in the VM queue for execution;\

Allocate $j$ to the PM with the lowest load and available capacity;\

Upload load $(CM)$ of the PM;} \

\nllabel{Prepartition Outer Loop End Line}
Compute CM of each PM and total partitions
\end{algorithm}

\textit{ Theorem 2: The computational complexity of Prepartition algorithm is $O(nlogm)$ using priority queue data structure where $n$ is the number of VM requests after pre-partition and $m$ is total number of PMs used}. \\
\noindent Proof: The priority queue is designed such that each element (PM) has a priority value (average capacity$\_$makespan), and each time the algorithm needs to select an element from it, the algorithm takes the one with the highest priority (the smaller average capacity$\_$makespan value is, the higher priority it is). Sorting $n$ numbers in a priority queue takes $O(n)$ time and a priority queue performs insertion and the extraction of minima in $O(logn)$ steps (detailed proof of the priority queue is shown in \cite{IEEEhowto:Kleinberg2}). Therefore, by using priority queue or related data structure, the algorithm can find a PM with the lowest average capacity$\_$makespan in $O(logm)$ time. Altogether, for $n$ requests, Prepartition algorithm has time complexity $O(nlogm)$.

\textit { Theorem 3: The approximation ratio of Prepartition algorithm is $(1+\epsilon)$ regarding the capacity\_makespan where $\epsilon$=$\frac{1}{k}$} and $k$ is the partition value (a preset constant).\\
Proof: This is because that each request has bounded capacity$\_$makespan by pre-parition based on ideal lower bound $P_0$. We sketch the proof as follows. Each job has start-time $s_i$, end-time $f_i$ and process time $p_i$=$f_i$-$s_i$. Consider the last job to finish (after scheduling all other jobs) and suppose this job starts at time $T_0$. All the machines must have been fully loaded up to capacity\_makespan $CM_0$, which gives $CM_0\leq$OPT. Since, for all jobs, we have $CM_i\leq\epsilon$ OPT (by the settting of Prepartition algorithm in equation (7)), this job finishes with load $CM_0$+$\epsilon$OPT. Hence, the schedule with capacity\_makespancan be no more than $CM_0$+$\epsilon$ OPT $\leq$ (1+$\epsilon$)OPT, this finishes the proof. \\
\subsection{Online Prepartition Algorithm}
For online VM allocations, scheduling decisions must be made without complete information about the entire job instances because jobs arrive one by one. We extend the offline Prepartition algorithm to online scenario as PrepartitionOn.

Let us consider there are $m$ PMs and  $L$ VMs (including the one just came) in a data center. Firstly define
\begin{equation}
B_d=min{\lceil\max_{1\le j\le L} {(CM_j)/2}, {\sum_{j=1}^{L} (CM_j)}/m\rceil}
\end{equation}
$B_d$ is called dynamic balance value, which is one half of the max capacity$\_$makespan of all current PMs or the ideal load balance value of all current PMs in the system, where $L$ is the number of VMs requests already arrived. Notice that the reason to set $B_d$ as one half of the max capacity$\_$makespan of all current PMs is to avoid large requests may cause imbalance in some cases.

Algorithm 4.2 shows the pseudo codes of PrepartitionOn algorithm. Since in online algorithm, the requests come one by one, the system can only capture the information of arrived requests.
When a new request comes into the system, the algorithm computes dynamic balance value by equation (8). To be noticed, $L$ represents the number of requests  already arrived,
and $m$ represents the number of PMs  in use. After the dynamic balanced value ($B_d$) is computed, then the initial request is
partitioned into several requests (segments) based on the partition value $k$. In these partitioned requests, the first one would be executed instantly, which will be allocated to
the PM with the lowest capacity\_makespan, while  others would be put back into the queue waiting to be executed. Then the algorithm picks up the next arrived request to follow
the
same partition and allocation process. After all requests are allocated, the algorithm computes the capacity$\_$makespan of each PM and find the total partition numbers for $n$
requests. Since the number of partitions and segments of each VM request are known at the moment of allocation, the system can prepare VM migration in advance so that process time
and
instability  of migration can be reduced.
\begin{algorithm}[!tb]
\SetArgSty{textnormal}
\caption{PrepartitionOn Algorithm}\label{PrepartitionOn }
\KwIn{VM requests come one by one indicated by their information (required VM type IDs, start times, ending times, requested capacity), $CM_i$ is the capacity$\_$makespan of request $i$}
\KwOut{Assign a PM ID to all requests and their partitions}
Initialization: set the partition value k, total partition number $P$=0;


\For {each arrived job $j$} {\nllabel{PrepartitonOn Outer Loop Begin Line}
Pick up the VM with the earliest start time in the VM queue to schedule\;
Compute $CM_j$ of VM $j$, and $B_d$ using Equ.~(8)\;

\If{$CM_j>\lceil(B_d/k)\rceil$} {partition $VM_j$ into multiple $\lceil(B_d/k)\rceil$ subintervals equally, consider each subinterval as a new request and add them into VM queue,~$P=P+\lceil \frac{CM_j}{B_d/k}\rceil$ \;
\nllabel{End If}
\Else{Allocate $j$ to PM with the lowest load and available capacity\;
Update load $(CM)$ of the PM\;}}
}\nllabel{PrepartitonOn Outer Loop End Line}
Compute $CM$ of each PM and output total number of partitions $P$\;
\end{algorithm}

%

\textit{ Theorem 4: The competitive ratio of PrepartitionOn is $(1+\frac{1}{k}-\frac{1}{mk})$} regarding the capacity\_makespan. \\
Proof:  Without loss of generality, we label PMs in order of non-decreasing final loads in PrepartitionOn. Denote $OPT$ and and $PrepartitionOn(I)$  respectively as the optimal load balance value of
corresponding offline scheduling
 and load balance value of PrepartitionOn for a given set of jobs $I$, respectively. Then the load of $PM_m$ defines the capacity$\_$makespan. The first $(m$-1) PMs each process a subset of the jobs and then experience a (possibly none) idle period. All PMs together finish a total capacity$\_$makespan
 $\sum_{i=1}^{n} CM_i$ during their busy periods. Consider the allocation of the last job  $j$to PM$_m$. By the scheduling rule of PrepartitionOn, PM$_m$ had the lowest load at the time
of allocation. Hence, any idle period on the first ($m$-1) PMs cannot be bigger than the capacity$\_$makespan of the last job allocated on PM$_m$ and hence cannot exceed the maximum
capacity$\_$makespan divided by $k$ (partition value), i.e., $\frac{\max_{1\leq i\leq n}CM_i}{k}$. We have \\
\begin{equation}
\footnotesize
m\times PrepartitionOn(I)\leq \sum_{i=1}^{n} CM_i +(m-1)\frac{\max_{1\leq i\leq n}CM_i}{k}
\end{equation}
which is equivalent to
\begin{equation}
\footnotesize
PrepartitionOn(I)\leq \frac{\sum_{i=1}^{n} CM_i}{m} +\frac{(m-1)\max_{1\leq i\leq n}CM_i}{mk}
\end{equation}
which is
\begin{equation}
 PrepartitionOn(I)\leq OPT+(\frac{1}{k}-\frac{1}{mk})OPT
\end{equation}
Note that  $\frac{\sum_{i=1}^{n} CM_i}{m}$ is the lower bound on $OPT(I)$ because the optimum capacity$\_$makespan cannot be smaller than the average capacity$\_$makespan on all PMs. And $OPT(I) \geq \max_{1\leq
i\leq n}CM_i$ since the largest job must be processed on a PM. We therefore have $PrepartitionOn(I)\leq (1+\frac{1}{k}-\frac{1}{mk})OPT$.

\textit{Theorem 5: The computational complexity of PrepartitionOn is $O(nlogm)$ using priority queue data structure, where $n$ is the number of VM requests after pre-partition and
$m$
is the total number of PMs used}. \\
\noindent Proof:   The  proof is exactly the same as in the proof for Theorem 2, we therefore omit it. \\

\section{Performance Evaluation}
\begin{table}
\footnotesize
\caption{8 types of virtual machines (VMs) in Amazon EC2}
\begin{center}
\begin{tabular}{|l|l|l||l|}
\hline Compute Units& Memory & Storage& VM Type
\\\hline
\hline 1 units & 1.875GB & 211.25GB & 1-1(1) \\
\hline 4 units & 7.5GB & 845GB & 1-2(2) \\
\hline 8 units & 15GB & 1690GB&1-3(3) \\
\hline 6.5 units& 17.1GB & 422.5GB &2-1(4)\\
\hline 13 units & 34.2GB & 845GB &2-2(5)\\
\hline 26 units & 68.4GB & 1690GB &2-3(6)\\
\hline 5 units & 1.875GB & 422.5GB &3-1(7) \\
\hline 20 units & 7GB & 1690GB &3-2(8)\\
\hline
\end{tabular} \\
\end{center}
\end{table}
\begin{table}
\caption{3 types of physical machines (PMs) suggested
}
\begin{center}
\footnotesize
\begin{tabular}{|l|l|l||l|}
\hline PM Pool Type& Compute Units & Memory& Storage
\\\hline
\hline Type 1 & 16 units & 30GB & 3380GB \\
\hline Type 2 & 52 units & 136.8GB & 3380GB \\
\hline Type 3 & 40 units & 14GB & 3380GB\\
\hline
\end{tabular} \\
\end{center}
\end{table}
In this part, we will present the simulation results between Prepartition algorithms and other existed algorithms. To achieve this goal, we used a Java simulator CloudSched (see Tian et al. \cite{IEEEhowto:Tian2}). For simulation, to be realistic and reasonable, we adopt data both from Normal distribution and Lawrence Livermore National Lab (LLNL) trace, see \cite{IEEEhowto:ESL} for detailed introduction about the trace.

All simulations are conducted on a computer configured with Intel i5 processor at 2.5GHz and 4GB memory. All VM requests are generated following Normal distribution. In offline algorithm comparisons, Round-Robin (RR) algorithm, Longest Process Time (LPT) algorithm and Post Migration Algorithm (PMG) are implemented.
\subsection{Offline Algorithm Performance Evaluation}
\begin{figure*}[!ht]
\begin{center}
{\includegraphics [width=1.0\textwidth,height=1.5in, angle=-0] {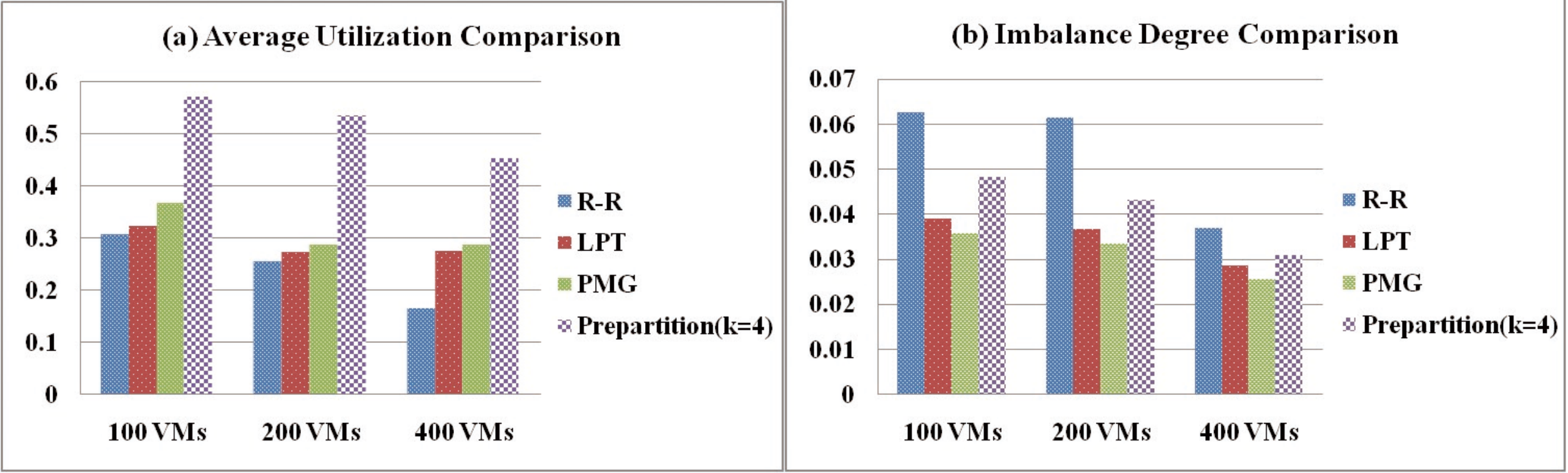}}
\caption{The offline algorithm comparison of average utilization (a) and imbalance degree (b) with LLNL trace}
\end{center}
\end{figure*}

\begin{figure*}[!ht]
\begin{center}
{\includegraphics [width=1.0 \textwidth, height=1.5in,angle=-0] {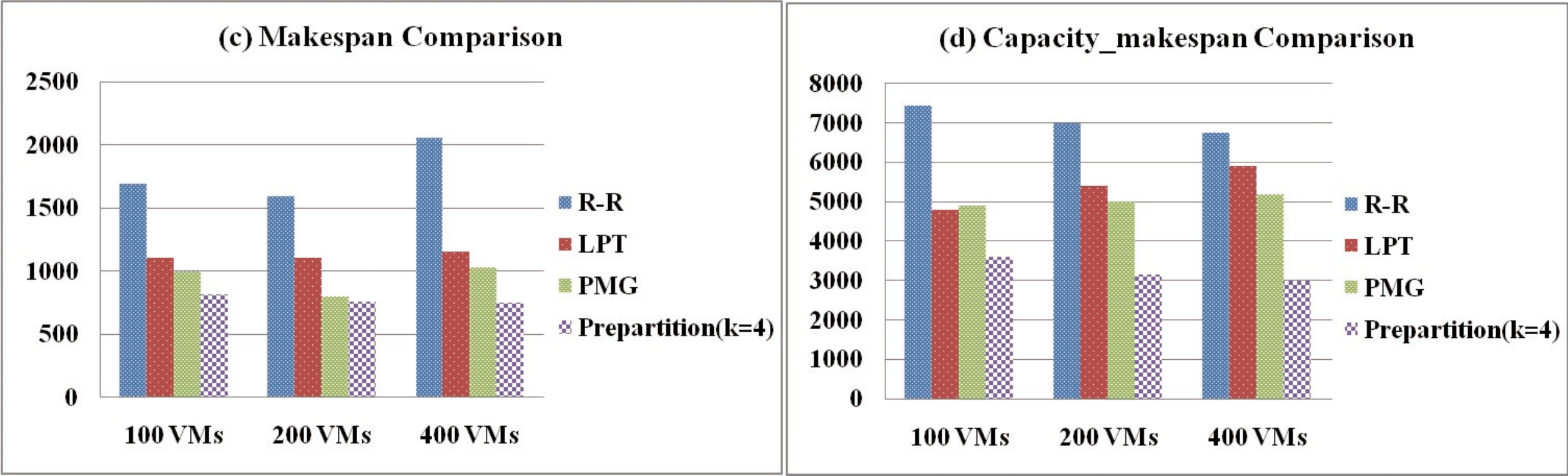}}
\caption{The offline algorithm comparison of makespan (c) and capacity\_makespan (d) with LLNL trace}
\end{center}
\end{figure*}
\noindent 1) Round-Robin Algorithm (R-R): a traditional load balancing scheduling algorithm by allocating the VM requests in turn to each PM that can provide required resource.\\
2) Longest Processing Time first (LPT): it sorts the VM requests by processing time in decreasing order firstly. Then allocating the requests in that order to the PM with the lowest load. In this paper, the lowest load means the lowest capacity$\_$makespan of all PMs.\\
3) Post Migration algorithm (PMG): Firstly, it processes the requests in the same way as LPT does. Then the average capacity$\_$makespan of all jobs is calculated. The up-threshold and low-threshold of the capacity$\_$makespan for the post migration are calculated through the average capacity$\_$makespan multiplied by a factor (in this paper we set the factor as 0.1, so the up-threshold is average capacity$\_$makespan multiplied by 1.1 and the low-threshold is multiplied by 0.9). Off course the factor can be set dynamically to meet different requirements; however, the larger the factor is, the higher imbalance is. A migration list is formed by collecting the VMs taken from PMs with capacity$\_$makespan higher than the low-threshold. The VMs would be taken from a PM only if the operation would not lead the capacity$\_$makespan of the PM to be less than the low threshold. After that, the VMs in the migration list would be re-allocated to a PM with capacity$\_$makespan less than the up-threshold. The VMs would be allocated to a new PM only if the operation would not lead  the capacity$\_$makespan of the PM to be higher than the up-threshold. There may be still some VMs left in the list, finally the algorithm allocates the left VMs to the PMs with the lowest capacity$\_$makespan until the list is empty. \\

In this paper, we adopt the Amazon EC2 configuration of VMs and PMs as shown in Table 1 and 2. Note that one compute unit (CU) has equivalent CPU capacity of a 1.0-1.2 GHz 2007 Opteron or 2007 Xeon processor \cite{IEEEhowto:Amazon}.

\textit{Observation 1. PMG is a best-effort trial heuristic for load balance. It does not guarantee a bounded or predefined load balance objective. This is validated in the following performance evaluation section.}

\subsubsection{Replay with LLNL Data Trace}\label{Sec:FirstModel}
As for realistic data, we adopt the log data at Lawrence Livermore National Lab (LLNL) \cite{IEEEhowto:ESL}. The log contains months of records collected by a large Linux cluster and has characteristics consistent with our problem model. Each line of data in that log file includes 18 elements, while we only need the request-ID, start-time, duration and number of processors (capacity demands) in our simulation. We convert the units from seconds in LLNL log file into minutes, because we set 5 minutes as a time slot length mentioned in previous section.

Fig.1 and Fig.2 show the average utilization, imbalance degree, makespan and capacity$\_$makespan comparison for different algorithms with LLNL data trace. From these figures, we can notice that Prepartition algorithm has better performance than other algorithms in average utilization, imbalance degree, makespan, capacity$\_$makespan. Prepartition algorithm has 10$\%$-20$\%$ higher average utilization than PMG and LPT, and 40$\%$-50$\%$ higher average utilization than Random-Robin (RR). Prepartition algorithm has 10$\%$-20$\%$ lower average makespan and capacity$\_$makespan than PMG and LPT, 5\% imbalance degree than LPT and 40$\%$-50$\%$ lower average makespan and capacity$\_$makespan than Random-Robin (RR).

With the partition value $k=4$, PMG algorithm has a quite similar value in imbalance value, so besides the above evaluations, we also vary the partition number $k$ from 4, 8 to 10 to compare the imbalance degree affects. In Figure 3, we can notice that larger $k$ value will induce a lower imbalance degree. Similarly, with a larger value, larger average utilization, lower makespan and capacity\_makespan can be acquired.

However, increasing the $k$ value will bring side-effects. The dominant one is running time, in Fig. 4, we compare the time costs under different partition value $k$, Prepartition algorithm with $k=8$ costs about 10\% more running time, and  with $k=10$ it costs 15\% more running time than Prepartition algorithm with $k=4$ on the average.  It is easy to understand that a larger $k$ value will produce a better load balance, which  leads to more partitions, and more partitions need more time to proceed. \\

\begin{figure}[!ht]
\begin{center}
{\includegraphics [width=0.5\textwidth, height=1.5in,angle=-0] {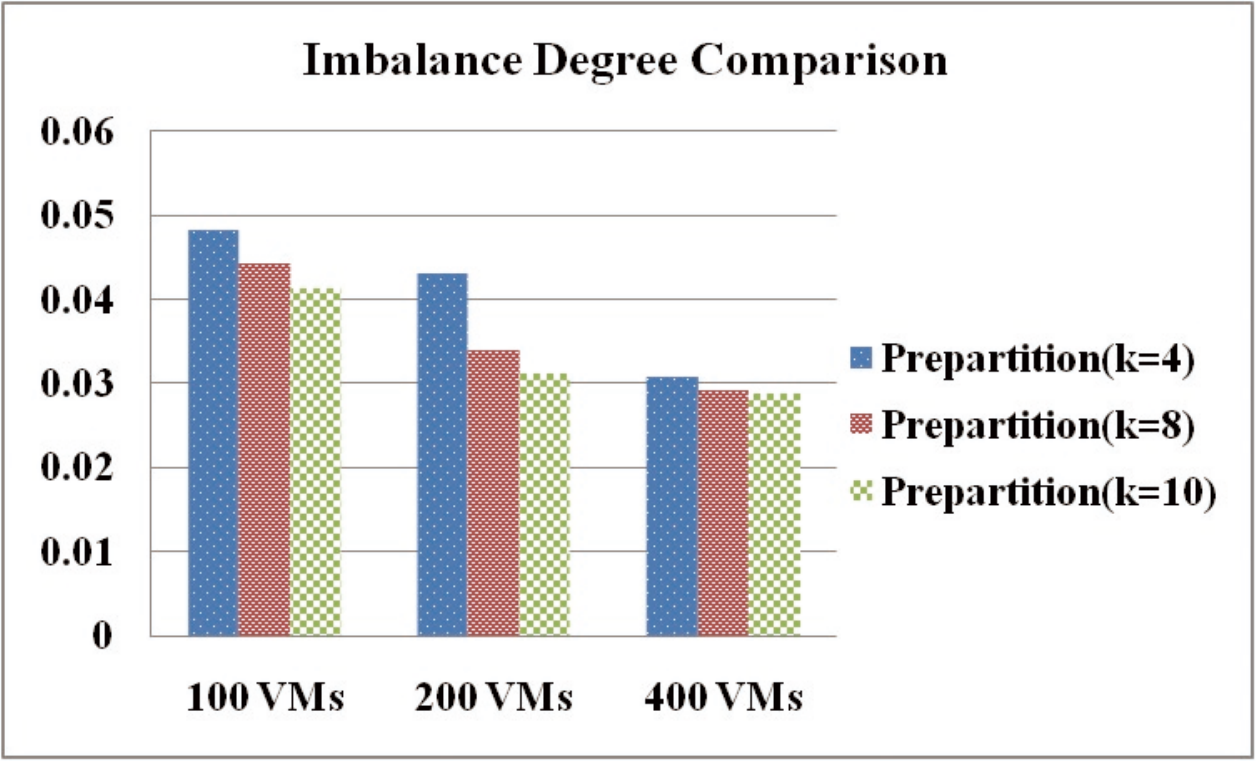}}
\caption{The comparison of imbalance value  by varying $k$ values}
\end{center}
\end{figure}

\begin{figure}[!ht]
\begin{center}
{\includegraphics [width=0.5\textwidth, height=1.5in,angle=-0] {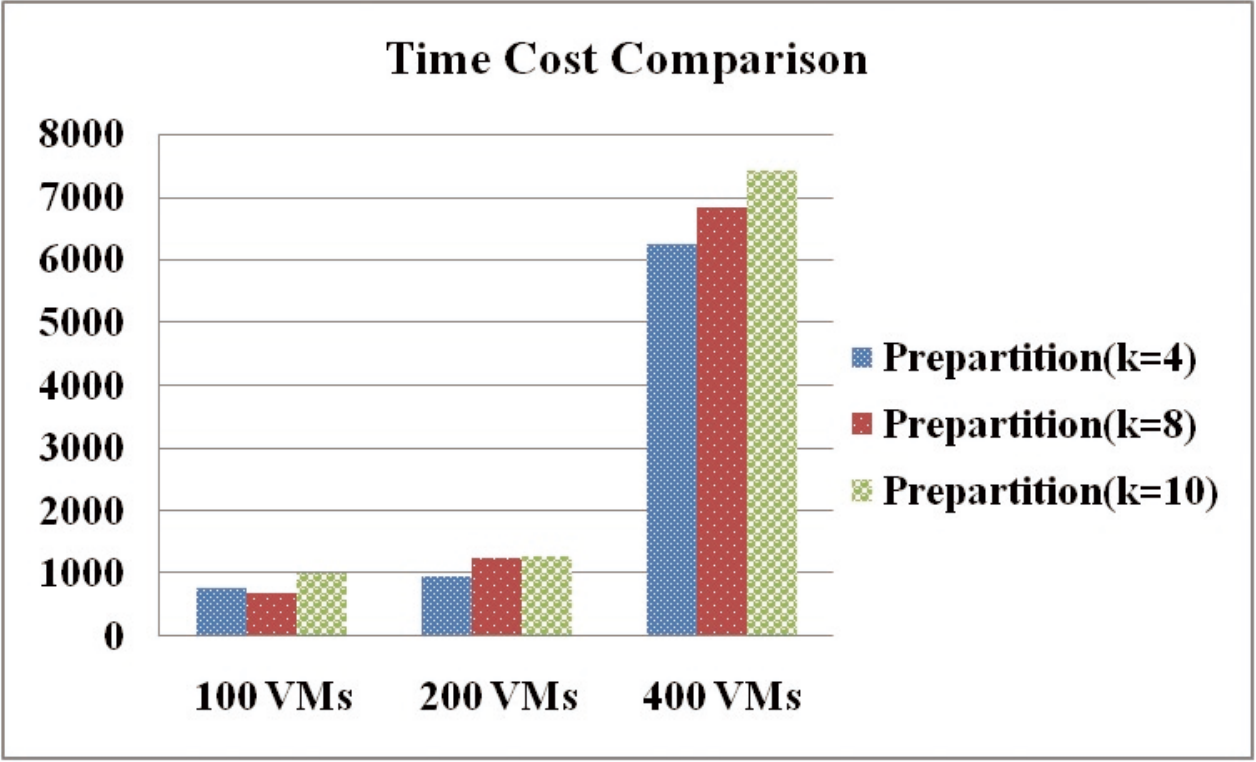}}
\caption{The comparison of time costs by varying $k$ values}
\end{center}
\end{figure}

\textit{Observation 2.  Whatever numbers of migrations to taken, post migration algorithm (PMG) just cannot achieve the same level of average utilization, makespan and capacity$\_$makespan as Prepartition does}.

This is because that Prepartition works in a much more refined and desired scale by prepartition based on reservation data while PMG is just a best-effort trial by migration.
\subsubsection{Results Comparison by Synthetic Data }\label{Sec:FirstModel}
We set 5 minutes as a slot, so 12 slots are for an hour, 288 slots are for a day. All requests satisfy the Normal distribution, with parameters mean $\mu$ and standard deviation $\delta$ as 864 (three days) and 288 (one day) respectively. After requests are generated in this way, we start the simulator to simulate the scheduling effects of different algorithms and comparison results are collected.
For collecting data, we firstly fix the $k$ value of Prepartition algorithm as 4; different types of VMs wit equal probabilities. Then we change the VMs numbers from $100$, $200$, $400$ and $1600$ to trace the tendency. Each set of data is the average values of 10-runs.
\begin{figure*}[ht]
\begin{center}
{\includegraphics [width=1.0\textwidth, height=1.5in,angle=-0] {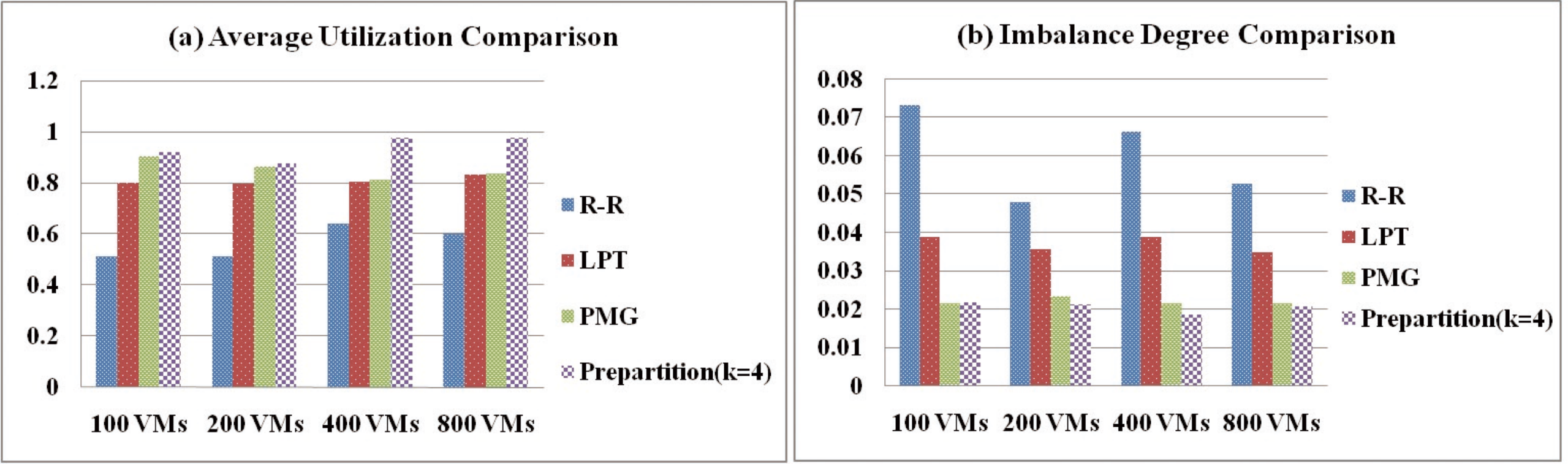}}
\caption{The offline algorithm comparison of average utilization (a) and imbalance degree (b) with Normal distribution }
\end{center}
\end{figure*}

\begin{figure*}[ht]
\begin{center}
{\includegraphics [width=1.0\textwidth, height=1.5in,angle=-0] {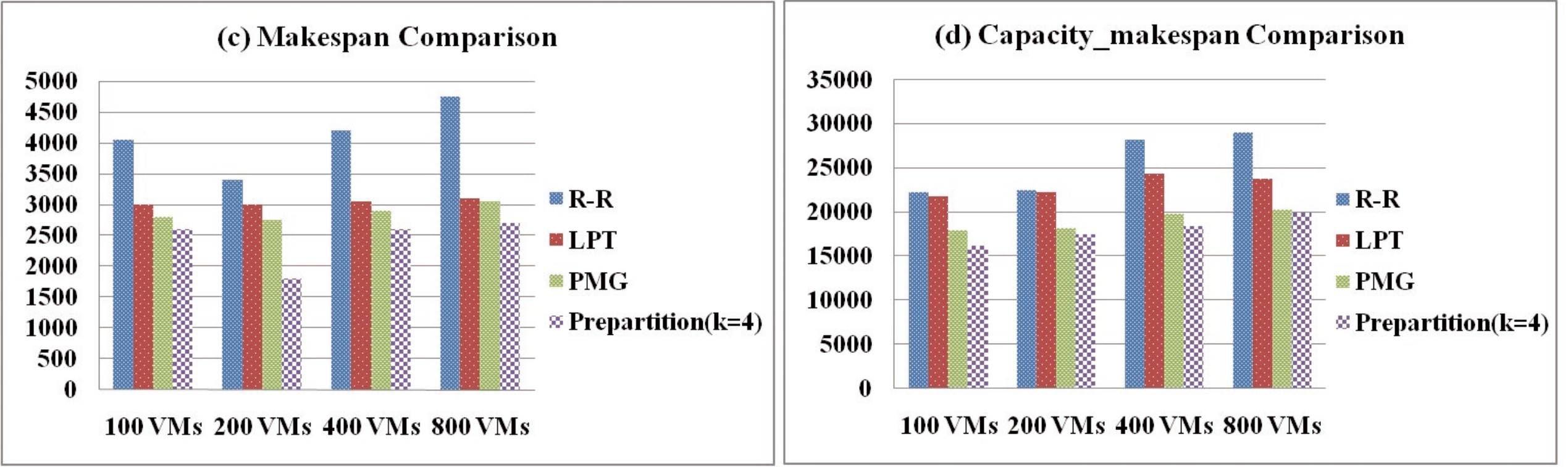}}
\caption{The offline algorithm comparison of makespan (c) and capacity\_makespan (d) with Normal distribution}
\end{center}
\end{figure*}
Fig.5 to Fig.6 show the average utilization, makespan and capacity$\_$makespan comparison of different algorithms respectively. From these figures, we can notice that Prepartition algorithm has 10$\%$-20$\%$ higher average utilization than PMG and LPT, and 40$\%$-50$\%$ higher average utilization than Random-Robin (RR); Prepartition algorithm has 8$\%$-13$\%$ lower average makespan and capacity$\_$makespan than PMG and LPT, and 40$\%$-50$\%$ lower average makespan and capacity$\_$makespan than Random-Robin (RR).
We can also notice that the PMG algorithm can improve the performance of LPT algorithm. LPT algorithm is better than R-R algorithm. Similar results are observed for the comparison of makespan.
The performance improvement of MIG algorithm is obtained from the extra migration operations. The VM migration enables a better load balance.

\subsection{Online Prepartition Algorithm}
In this part, we will present the simulation results between PrepartitionOn algorithm and other three existed algorithms.
Random, Round-Robin, Online Resource Scheduling Algorithm (OLRSA) \cite{IEEEhowto: Xu} and PrepartitionOn Algorithm are implemented to compare:\\
1) Random Algorithm: a scheduling algorithm that randomly allocates the requests to a PM that can provide required resource.\\
2) Round-Robin Algorithm(R-R): a traditional load balancing scheduling algorithm by allocating the VM requests in turn to each PM that can provide required resource.\\
3) OLRSA algorithm: an online scheduling algorithm, it computes the capacity$\_$makespan of each PM and sort the PM by capacity$\_$makespan in descending order. This algorithm always allocates the request to the PM  with
the lowest capacity\_makespan and required resource.\\

\subsubsection{Replay with LLNL Data Trace}
\begin{figure*}[!ht]
\begin{center}
{\includegraphics [width=1.0\textwidth, height=1.5in,angle=-0] {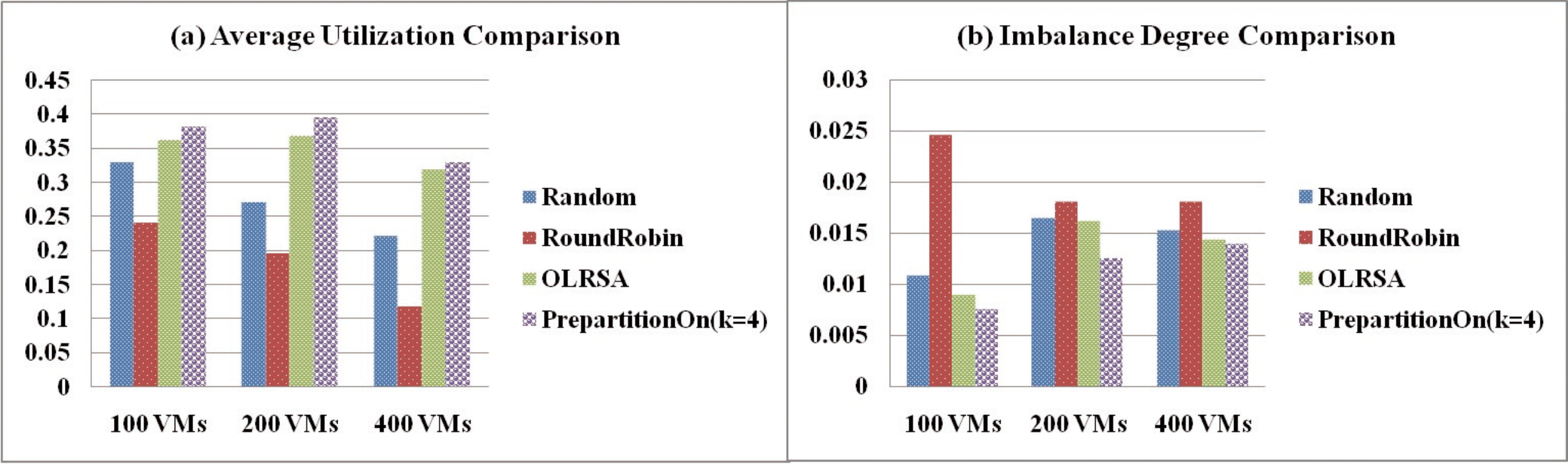}}
\caption{The  online algorithm comparison of average utilization (a) and imbalance degree (b) with LLNL trace}
\end{center}
\end{figure*}

\begin{figure*}[!ht]
\begin{center}
{\includegraphics [width=1.0\textwidth, height=1.5in,angle=-0] {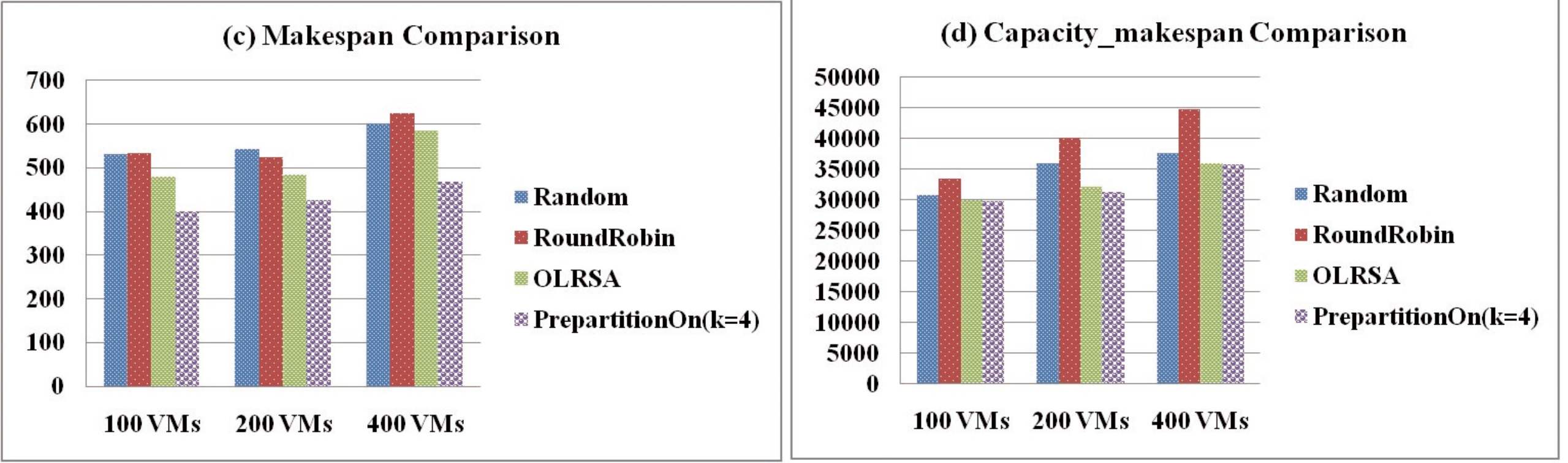}}
\caption{The online algorithm comparison of makespan (c) and capacity\_makespan (d) with LLNL trace}
\end{center}
\end{figure*}
For realistic data, we utilize the log data at Lawrence Livermore National Lab (LLNL)\cite{IEEEhowto:ESL} because the data is suitable for our research problem.
Fig. 7 to Fig. 8 illustrate the comparisons of  the average utilization, imbalance degree, makespan, capacity\_makespan. From these figures, we can notice that PrepartitionOn  shows
the highest average utilization, lowest imbalance degree, and lowest makespan. As for capacity\_makespan, OLRSA has been proved
much better performance compared with random and round-robin algorithms, and PrepartitionOn still improves 10$\%$-15$\%$ in average utilization, 20$\%$-30$\%$ in imbalance degree,
and
5$\%$ to 20$\%$ in makespan than OLRSA.
\subsubsection{Results Comparison by Synthetic Data}

\begin{figure*}[!ht]
\begin{center}
{\includegraphics [width=1.0\textwidth, height=1.5in,angle=-0] {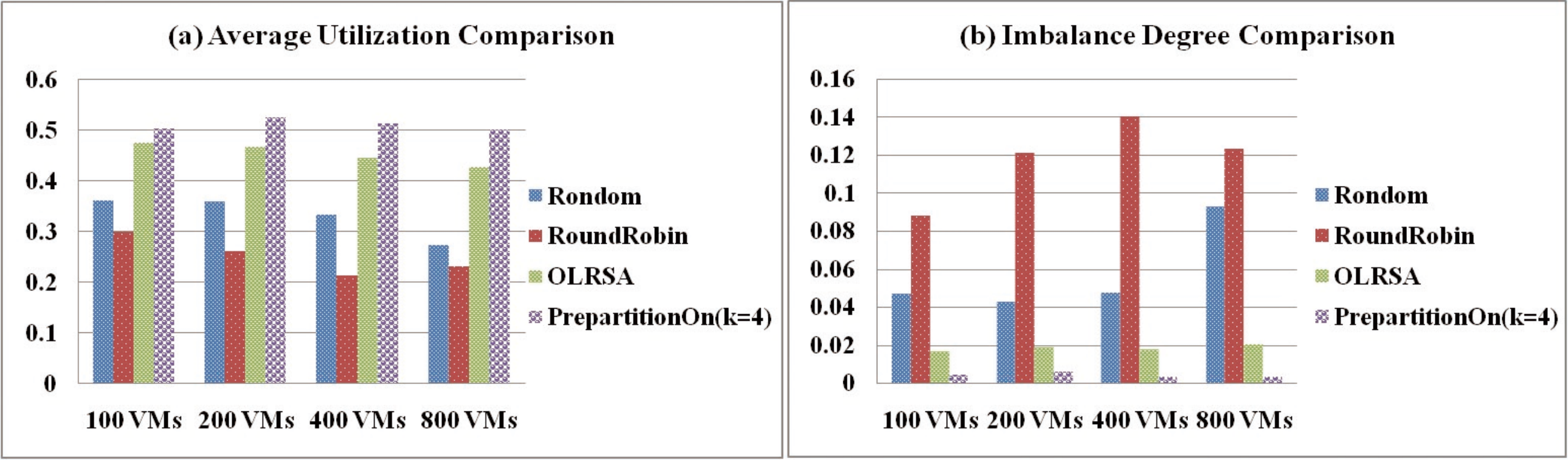}}
\caption{The online algorithm comparison of average utilization (a) and imbalance degree (b) with Normal distribution}
\end{center}
\end{figure*}

\begin{figure*}[!ht]
\begin{center}
{\includegraphics [width=1.0\textwidth, height=1.5in,angle=-0] {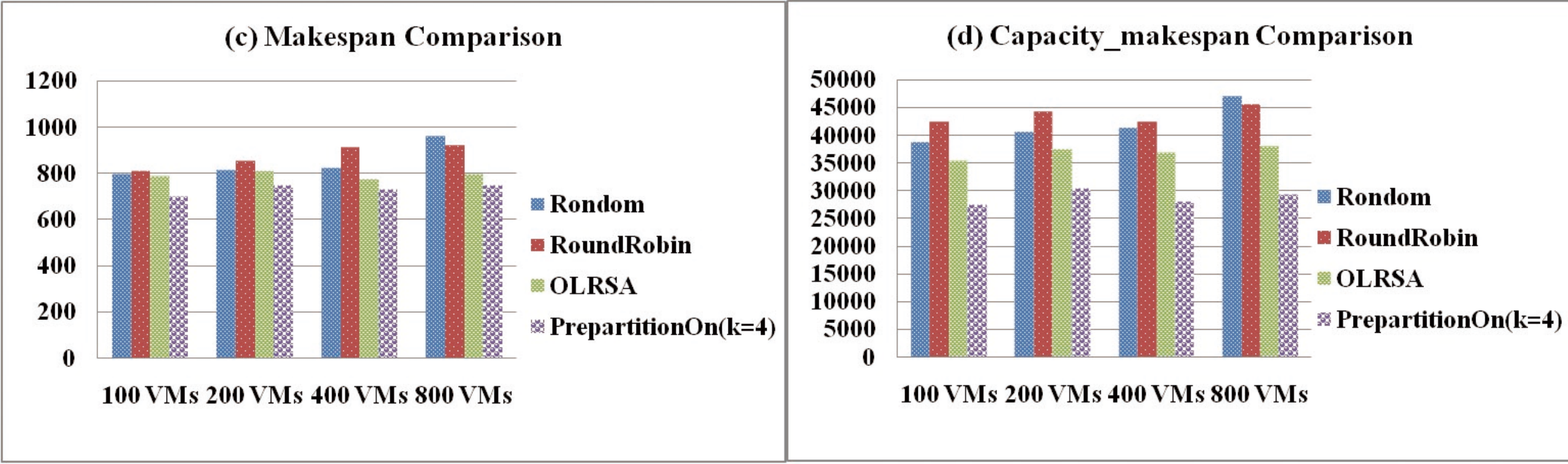}}
\caption{The online algorithm comparison of makespan (c) and capacity\_makespan (d) with Normal distribution}
\end{center}
\end{figure*}
We set 5 minutes as a slot, so 12 slots are for an hour, 288 slots are for a day. All requests satisfy the Normal distribution, with parameters mean $\mu$ and standard deviation
$\delta$ as 864 (3 days) and 288 (1 day). We set that different types of VMs have equal probabilities, then we change the requests generation approach to produce different size of requests to
trace the tendency. From Fig. 9 to 10, we can see that PrepartitionOn  has better performance in average utilization, imbalance degree, makespan and capacity\_makespan. Comparing to
OLRSA, PrepartitionOn  still improves about 10$\%$ in average utilization, 30$\%$-40$\%$ in imbalance degree, 10$\%$-20$\%$ in makespan, as well as 10$\%$-20$\%$ in
capacity\_makespan.

It is apparent that large $k$ values may bring side effects since it will need more number of partitions. In Fig. 11, we compare the time costs (simulated  with LLNL data and the time unit is mini second) under different partition value $k$, PrepartitionOn algorithm with $k=3$ takes about 10\% less running time than that with $k$=4, and  $k=2$ takes 15\% less running time than that with $k=4$. It is easy to understand that a larger $k$ value will produce a better load balance with longer process time.  We also observe that larger $k$ value will induce a lower capacity\_makespan value. Similarly, with a larger $k$ value, larger average utilization, lower imbalance degree and makespan are obtained. \\
\begin{figure}[!ht]
\begin{center}
{\includegraphics [width=.45\textwidth, height=1.5in,angle=-0] {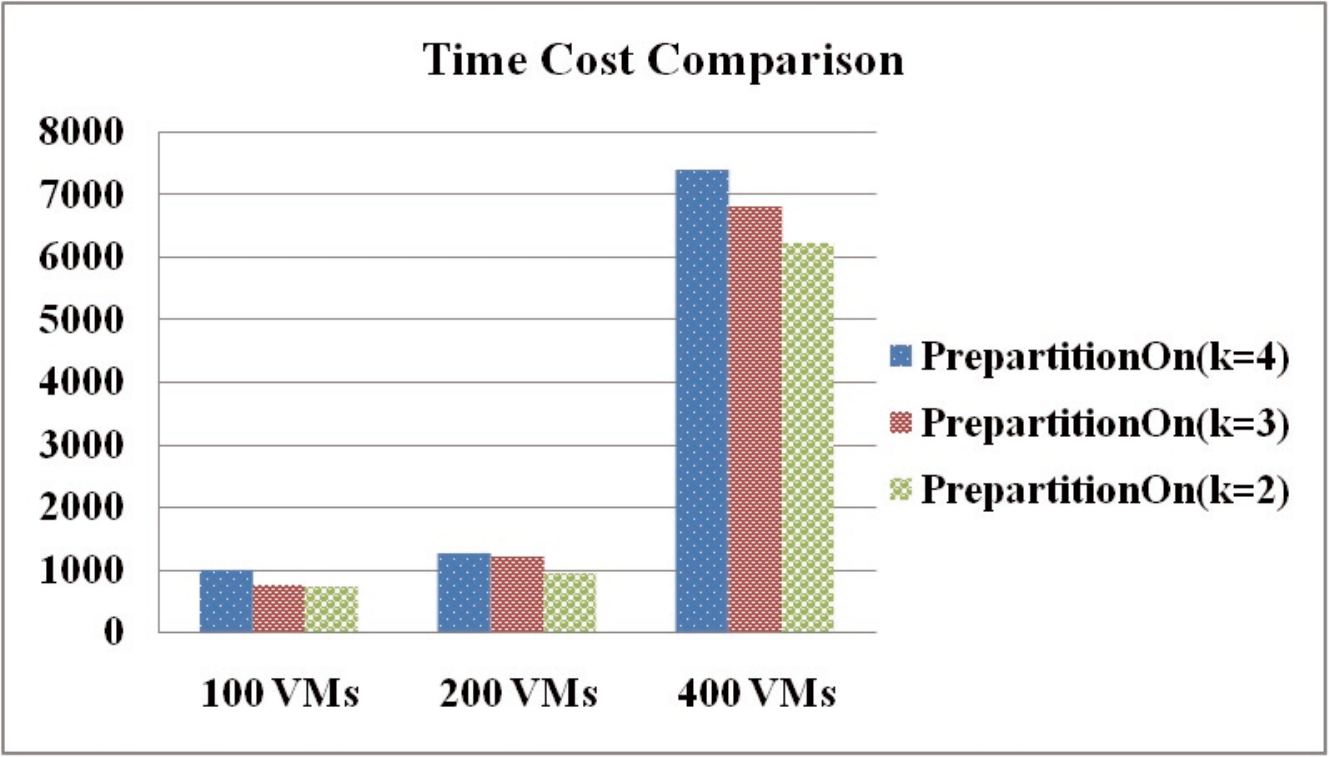}}
\caption{The comparison of time costs for PrepartitionOn by varying $k$ values}
\end{center}
\end{figure}

\section{conclusion}
In this paper, to reflect the feature of capacity sharing and fixed interval constraint of VM scheduling in Cloud data centers, we propose new offline and online load balancing algorithms. Theoretically we prove that offline Prepartition is a (1+$\epsilon$)-approximation where $\epsilon$=$\frac{1}{k}$ and $k$ is a positive integer. By increasing $k$ it is possible to be very close to optimal solution, i.e.,  by setting $k$ value, it is also possible to achieve predefined load balance goal as desired because offline Prepartition is a (1+$\frac{1}{k}$)-approximation and online Prepartition (PrepartitionOn) has competitive ratio $(1+\frac{1}{k}-\frac{1}{mk})$. Both synthetic and trace driven simulations have validated theoretical observations and shown Prepartition algorithm has better performance than a few existing algorithms at average utilization, imbalance degree, makespan, and capacity$\_$makespan both for offline and online algorithms.
There are still a few research issues can be considered:
\begin{itemize}
\item
making suitable choice between total partition numbers and load balance objective. Prepartition algorithm can achieve desired load balance objective by setting suitable $k$ value. It may need large number of partitions so that the number of migrations can be large depending on the characteristics of VM requests. For example in EC2 \cite{IEEEhowto:Amazon}, the duration of VM reservations varies from a few hours to a few months, we can classify different types of VMs based on their durations (capacity$\_$makespans) firstly, then applying Prepartition will not have large partition number for each type.
In practice we need analyzing traffic patterns to make the number of partitions (premigrations) reasonable so that the total costs, including running time and migrations,  are not very high.
\item
considering heterogeneous configuration of PMs and VMs. We mainly consider that a VM requires a portion of total capacity from a PM. This is also applied in EC2 and Knauth et al. \cite{Knauth2012}. When this is not true, multi-dimensional resources such as CPU, memory and bandwidth etc. have to be considered together or separately in the load balance, see \cite{IEEEhowto:Singh} and \cite{IEEEhowto:Sun} for a detailed discussion about considering multi-dimensional resources.
\item
Considering precedence constraints among different VM requests. In reality, some of VM reservations may be more important than others, we should extend current algorithm to consider this case.

\end{itemize}

\section*{Acknowledgement}
This research is partially supported by China National Science Foundation (CNSF) with project ID 61450110440.  Partial results, especially the offline scheduling algorithm of this paper was presented in the conference of ICC 2014, Sydney, Australia\cite{IEEEhowto:Tian5}.

\end{document}